\documentclass[twocolumn,english,floatfix]{revtex4-2}
\usepackage[T1]{fontenc}
\usepackage{color}
\usepackage{amsmath}
\usepackage{graphicx}
\usepackage{braket}
\usepackage{bbold}
\usepackage{natbib}
\usepackage{adjustbox}
\usepackage{tabularx}

\usepackage{multirow}
\usepackage{siunitx}
\usepackage[utf8]{inputenc}
\usepackage[dvipsnames]{xcolor}

{
\makeatletter
\def\frontmatter@thefootnote{%
 \altaffilletter@sw{\@fnsymbol}{\@fnsymbol}{\csname c@\@mpfn\endcsname}%
}%
\makeatother

\newcommand{\gstate}{\ket{\text{0}}}

\newcommand{\estate}{\ket{\text{1}}}
\newcommand{\fstate}{\ket{\text{2}}}

\renewcommand{\eqref}[1]{\mbox{Eq.~(\ref{#1})}}

\newcommand{\figref}[1]{\mbox{Fig.~\ref{#1}}}
\newcommand{\figpanel}[2]{Fig.~\hyperref[#1]{\ref*{#1}(#2)}}
\newcommand{\figpanels}[3]{Fig.~\hyperref[#1]{\ref*{#1}(#2)-(#3)}}
\newcommand{\figpanelNoPrefix}[2]{\hyperref[#1]{\ref*{#1}(#2)}}

\usepackage[colorlinks=true,urlcolor=blue,citecolor=blue,linkcolor=blue]{hyperref}

\makeatother

\usepackage{babel}
\usepackage[displaymath, mathlines]{lineno}

\AtBeginDocument{%
  \setlength{\linenumbersep}{\dimexpr\oddsidemargin+0.45in-\linenumberwidth\relax}%
}

\begin{document}
	
	\title{Fast unconditional reset and leakage reduction in fixed-frequency transmon qubits}
	\def\andname{\hspace*{-0.5em},}
	\author{Liangyu Chen$^{1}$}
	\email[email: ]{liangyuc@chalmers.se}
	\email[\\email: ]{jonas.bylander@chalmers.se}
	\email[\\email: ]{tancredi@chalmers.se }
	\author{Simon Pettersson Fors$^{1}$, Zixian Yan$^{1}$, Anaida Ali$^{1}$, Tahereh Abad$^{1}$, Amr Osman$^{1}$, Eleftherios Moschandreou$^{1}$, Benjamin Lienhard$^{2,3}$, Sandoko Kosen$^{1}$, Hang-Xi Li$^{1}$, Daryoush Shiri$^{1}$, Tong Liu$^{1}$, Stefan Hill$^{1}$,  Abdullah-Al Amin$^{1}$, Robert Rehammar$^{1}$, Mamta Dahiya$^{1}$, Andreas Nylander$^{1}$, Marcus Rommel$^{1}$, Anita Fadavi Roudsari$^{1}$, Marco Caputo$^{4}$, Leif Grönberg$^{4}$, Joonas Govenius$^{4}$, Miroslav Dobsicek$^{1}$, Michele Faucci Giannelli$^{1}$, Anton Frisk Kockum$^{1}$, Jonas Bylander$^{1}$, Giovanna Tancredi$^{1,*}$}

	\affiliation{$^{1}$Department of Microtechnology and Nanoscience, Chalmers University of Technology, 41296 Gothenburg, Sweden.\\
	$^{2}$Department of Chemistry, Princeton University, Princeton, NJ 08544, USA\\
	$^{3}$Department of Electrical and Computer Engineering, Princeton University, Princeton, NJ 08544, USA\\
	$^{4}$VTT Technical Research Centre of Finland, FI-02044 VTT, Finland}
	
	\begin{abstract}
    The realization of fault-tolerant quantum computing requires the execution of quantum error-correction (QEC) schemes, to mitigate the fragile nature of qubits.
    In this context, to ensure the success of QEC, a protocol capable of implementing both qubit reset and leakage reduction is highly desirable. We demonstrate such a protocol in an architecture consisting of fixed-frequency transmon qubits pair-wise coupled via tunable couplers -- an architecture that is compatible with the surface code. We use tunable couplers to transfer any undesired qubit excitation to the readout resonator of the qubit, from which this excitation decays into the feedline. 
    In total, the combination of qubit reset, leakage reduction, and coupler reset takes only 83~\unit{\nano\second} to complete. 
    Our reset scheme is fast, unconditional, and achieves fidelities well above 99\,\%, thus enabling fixed-frequency qubit architectures as future implementations of fault-tolerant quantum computers. 
    Our protocol also provides a means to both reduce QEC cycle runtime and improve algorithmic fidelity on quantum computers.
	\end{abstract}
	
	\date{\today}
	
	\maketitle
	
\section*{Introduction}
	
 

    
    


To achieve fault-tolerant quantum computing, quantum error correction (QEC) algorithms are crucial components in preserving the quantum information in a logical qubit. Among QEC algorithms, the surface code has been shown to be a promising platform in recent implementation with superconducting qubits\cite{Google2021, krinner2021realizing,zhao2022realization,Google2022,Google2024}. In the surface code, the physical qubits are designated to be either data or ancilla qubits in a checkerboard pattern \cite{Preskill1997,Fowler2012, Martinis2015}. The data qubits store the quantum information, and are parity-checked by measuring the ancilla qubits in each error-correction cycle. Each parity check is followed by a reset operation on the ancilla qubits to return them to the ground state and prepare them for the next round of error detection. However, during the cycle, data and ancilla qubits are prone to leak outside the computational subspace, thus fatally compromising the error-correction algorithms \cite{GoogleLeakage2021,Marques2023,Miao2023,Battistel2021,Yang2024}. Therefore, both a fast and high-fidelity reset for ancilla qubits and a leakage-reduction unit (LRU) for data qubits are instrumental for practical error correction.\par



In general, qubit reset is used to speed up the algorithm runtime as the qubit lifetimes improve and the waiting time for the qubit excitation to naturally decay is significant \cite{Egger2018,Geerlings2013}. Qubit-reset protocols can be either active or passive depending on whether or not they require knowledge of the qubit state. In active reset, qubits are reset conditioned on the measured results \cite{Moreira2023}. If the qubit population is in the first excited state $\estate$, a $\pi$-pulse is sent to the qubit to drive the population back to the ground state $\gstate$ via real-time feedback. The primary limitation for active reset is the feedback time of the control electronics, and the success rate of the feedback operation depends on the readout fidelity. Passive reset is unconditional; it depopulates the qubit excited-state population regardless of its initial state \cite{Egger2018, Geerlings2013,GoogleLeakage2021, Zhou2021, Magnard2018, Yang2024}. Existing passive reset schemes typically require additional drive signals, flux-tunable qubits, or additional control elements on the device. These challenges lead to increased difficulties when scaling up to a larger number of qubits. 

Implementing an LRU requires resetting only the $\fstate$-state or higher-energy states without disturbing the computational subspace. An LRU can be implemented either by directly coupling the $\fstate$-state population to a lossy resonator \cite{Marques2023,Battistel2021, lacroix2023} or by performing a SWAP gate to transfer the $\fstate$-state population to another element on the processor \cite{Miao2023,Yang2024}. It is crucial to develop an LRU that complements the chosen reset strategy. Although LRUs are mainly discussed within the context of QEC, they can be applied during any quantum algorithm to reduce the accumulation of leakage errors.\par


\begin{figure}[t!]
\centering
 \includegraphics[width=3.3in, keepaspectratio=true]{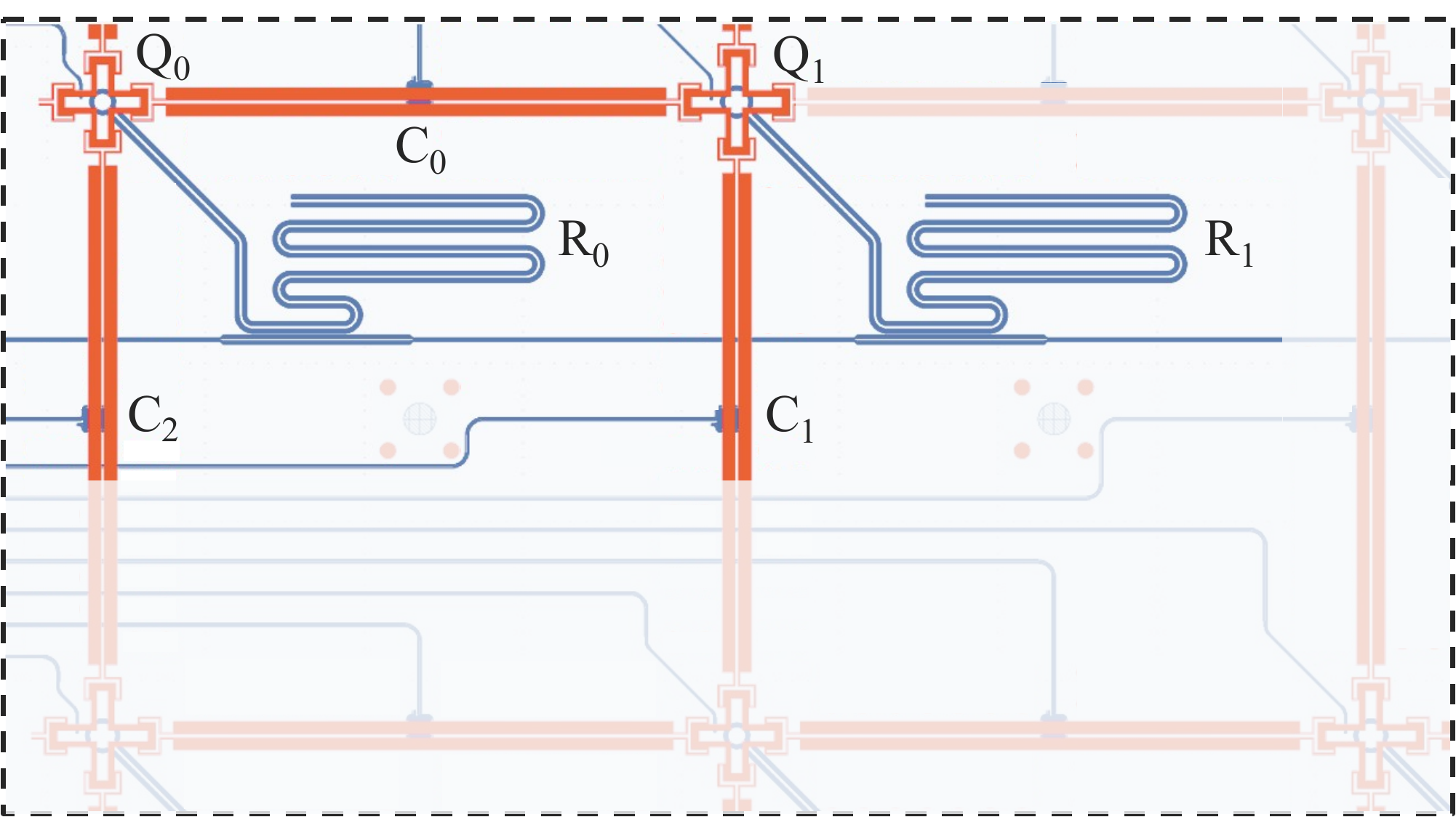}
 \caption {\textbf{Subset of the 25-chip device under test.} A pair of qubits, with their respective readout resonators and three of their surrounding couplers, are involved in the experiments. Qubit $\text{Q}_{0}$ ($\text{Q}_{1}$) is treated as the ancilla (data) qubit. Couplers $\text{C}_{0}$ and $\text{C}_{2}$ are utilized to reset qubit $\text{Q}_{0}$, and the leakage reduction unit on qubit $\text{Q}_{1}$ is implemented via coupler $\text{C}_{1}$. The device is designed and fabricated as a flip-chip architecture \cite{Kosen2022}, with the elements in red placed on the qubit chip and those in blue on the control chip interposer.}
\label{fig:device}
\end{figure}


In this article, we demonstrate both high-fidelity passive qubit reset and leakage reduction in an architecture using fixed-frequency transmon qubits \cite{Koch2007} pair-wise coupled via tunable couplers \cite{Gambetta2016, Yan2018, Kosen2022}. We propose a reset and leakage reduction scheme that utilizes the tunable coupler as means to transfer excitations from a qubit to its readout resonator, from which the excitations can then decay into the feedline. We achieve five objectives with our schemes: 1) a fast semi-adiabatic $\estate$- to $\gstate$-state reset with an error of $(1.87\pm1.12)\times10^{-3}$ within 9~\unit{\nano\second}, 2) an adiabatic reset that can also move both the $\estate$- and $\fstate$-state populations to the $\gstate$ state at the same time, reaching a $(7.87\pm1.94)\times10^{-3}$ error within 61~\unit{\nano\second}, 3) an LRU that removes the $\fstate$-state population back to the $\estate$ state without disrupting the computational subspace of the data qubits, with an error of $9.50\times10^{-3}$ in 5~\unit{\nano\second}, 4) a coupler reset unit that dissipates the excitation through the qubit readout resonator within 22~\unit{\nano\second}, and 5) simultaneous reset of ancilla qubits and leakage reduction of the data qubits in a total time of 83~\unit{\nano\second}. All operations have negligible impact on the other qubits on the chip and are easy to implement with the current fixed-frequency qubits and tunable-coupler architecture without the need for any additional hardware resources.\par

\begin{figure}[t!]
\centering
  \includegraphics[width=3.3in, keepaspectratio=true]{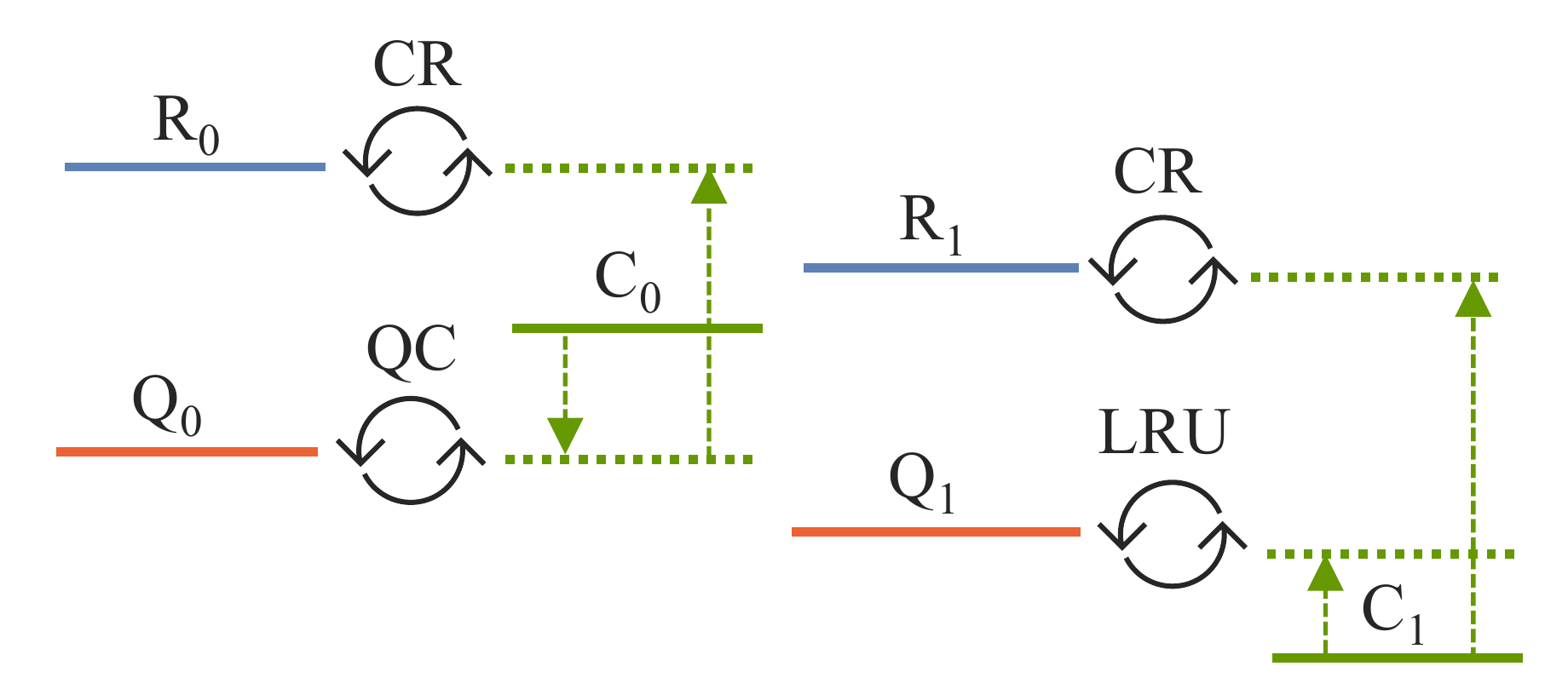}
 \caption {\textbf{Schematic of the reset and leakage-reduction protocol with energy levels of the qubit-resonator-coupler subsystem.} In a pair-wise coupled system, two fixed-frequency qubits, $\text{Q}_{0}$ and $\text{Q}_{1}$, are coupled via a frequency-tunable coupler $\text{C}_{0}$ controlled by an applied magnetic flux. Each qubit is also coupled to a dedicated readout resonator, denoted as $\text{R}_{0}$ and $\text{R}_{1}$. In the reset protocol, the coupler $\text{C}_{0}$ is first tuned on resonance with $\text{Q}_{0}$ to implement a qubit-coupler (QC) SWAP gate. $\text{Q}_{1}$ is not affected due to the relatively large detuning between the two qubits. For the leakage reduction unit (LRU), coupler $\text{C}_{1}$ is used, initially biased below the $\text{Q}_{1}$ frequency. Afterwards, both couplers are tuned to higher frequencies to interact with the resonators of either qubit to implement the coupler-resonator (CR) SWAP gate. The excitation will then decay into the environment via the readout feedline.}
\label{fig:energy}
\end{figure}

The experimental demonstration is carried out in a two-qubit subset of a 25-qubit device, which is illustrated in \figref{fig:device}. More details on the device, fabrication, and performance are provided in the Refs. \cite{Kosen2022,Kosen2024}. The qubit s are designed to be in two frequency groups in a checkerboard pattern, with neighbouring qubits separated in frequency by roughly 600~\unit{\mega\hertz} \cite{Osman2023}. The subset consists of two fixed-frequency transmon qubits $\text{Q}_{0}$ and $\text{Q}_{1}$ \cite{Koch2007} with transition frequencies $\omega_{Q_{i}}/2\pi$ at $5.176$ and $4.534$~\unit{\giga\hertz}, and anharmonicities of $\alpha_{Q_{i}}/2\pi$ at $-256$ and $-158$~\unit{\mega\hertz} for $i = 0$ and $1$, respectively. Each qubit is coupled with a strength $g_{i}$ to a readout resonator $\text{R}_{i}$ of frequency $\omega_{R_{i}}/2\pi = 6.752$ and $6.308$~\unit{\giga\hertz} for $i = 0$ and $1$, respectively. Both resonators are coupled to the same feedline. There is a tunable coupler between each pair of qubits arranged in a square grid on the chip. In this experiment, three of the couplers ($\text{C}_{0}$, $\text{C}_{1}$, and $\text{C}_{2}$) surrounding the pair of qubits are used. The qubit-to-coupler coupling rates $g_{QC}/2\pi$ are 40~\unit{\mega\hertz} (60~\unit{\mega\hertz}) for qubit $\text{Q}_{0}$ ($\text{Q}_{1}$), while the qubit-resonator coupling rates $g_{QR}/2\pi$ are around 50~\unit{\mega\hertz}. The coupling between the coupler and resonator has two main sources: a direct capacitive coupling due to their proximity on the chip and an indirect coupling mediated by the qubit. The device is cooled down to 10~mK and a microwave setup is used to measure the signal transmitted through the feedline. The complete experimental setup and device parameters resulting from the basic characterization are reported in Section I of the Supplementary Note. \par

The energy levels of the two-qubit subsystem in the single-excitation manifold are sketched out in \figref{fig:energy}. The high-frequency qubit $\text{Q}_{0}$ is treated as the ancilla qubit that we aim to reset, and the low-frequency qubit $\text{Q}_{1}$ acts as the data qubit that needs leakage reduction. To implement the reset protocol on the ancilla qubit, the first step is to do a qubit–coupler (QC) SWAP gate where the excitation is moved from qubit $\text{Q}_{0}$ to coupler $\text{C}_{0}$. $\text{C}_{0}$ is parked above $\text{Q}_{0}$ to avoid any interaction with $\text{Q}_{1}$. Using a similar mechanism, the LRU is implemented on qubit $\text{Q}_{1}$ via coupler $\text{C}_{1}$, parked below $\text{Q}_{1}$. The QC SWAP and the LRU can be excuted simultaneously on the respective qubit. Afterwards, the last step of the protocol involves the coupler–resonator (CR) SWAP gates for all couplers in parallel. The resonator excitation can then be dissipated through the feedline.\par

\section*{RESULTS}

\subsection*{Qubit reset with adiabatic pulses}
For qubit reset, one main scenario to consider is the ability to reset not just the qubit $\estate$-state population, but also that of the higher energy states, due to the accumulation of leakage population outside of the computational subspace. For this reason, an adiabatic pulse is chosen to perform qubit reset due to the possibility of transferring the population from multiple energy levels simultaneously. The pulse shape is calculated from the instantaneous approximate adiabatic condition. Given a Hamiltonian $H(t)$, for any two adjacent eigenstates $\ket{\Psi_n(t)}$ and $\ket{\Psi_m(t)}$ and corresponding eigenvalues $E_n(t)$ and $E_m(t)$, the evolution with duration $\tau$ is approximately adiabatic if the following condition is met:\par
\begin{equation}
\frac{\max\limits_{0\leq t \leq \tau}{\left|\bra{\Psi_n(t)} \frac{\partial{H(t)}}{\partial t}\ket{\Psi_m(t)} \right|}}{\max\limits_{0\leq t \leq \tau} |E_n(t) - E_m(t)|^2} \ll 1.
\label{eqn:global_adiabatic_condition}
\end{equation} 
\begin{figure}[t!]
\centering
\includegraphics[width=3.3in, keepaspectratio=true]{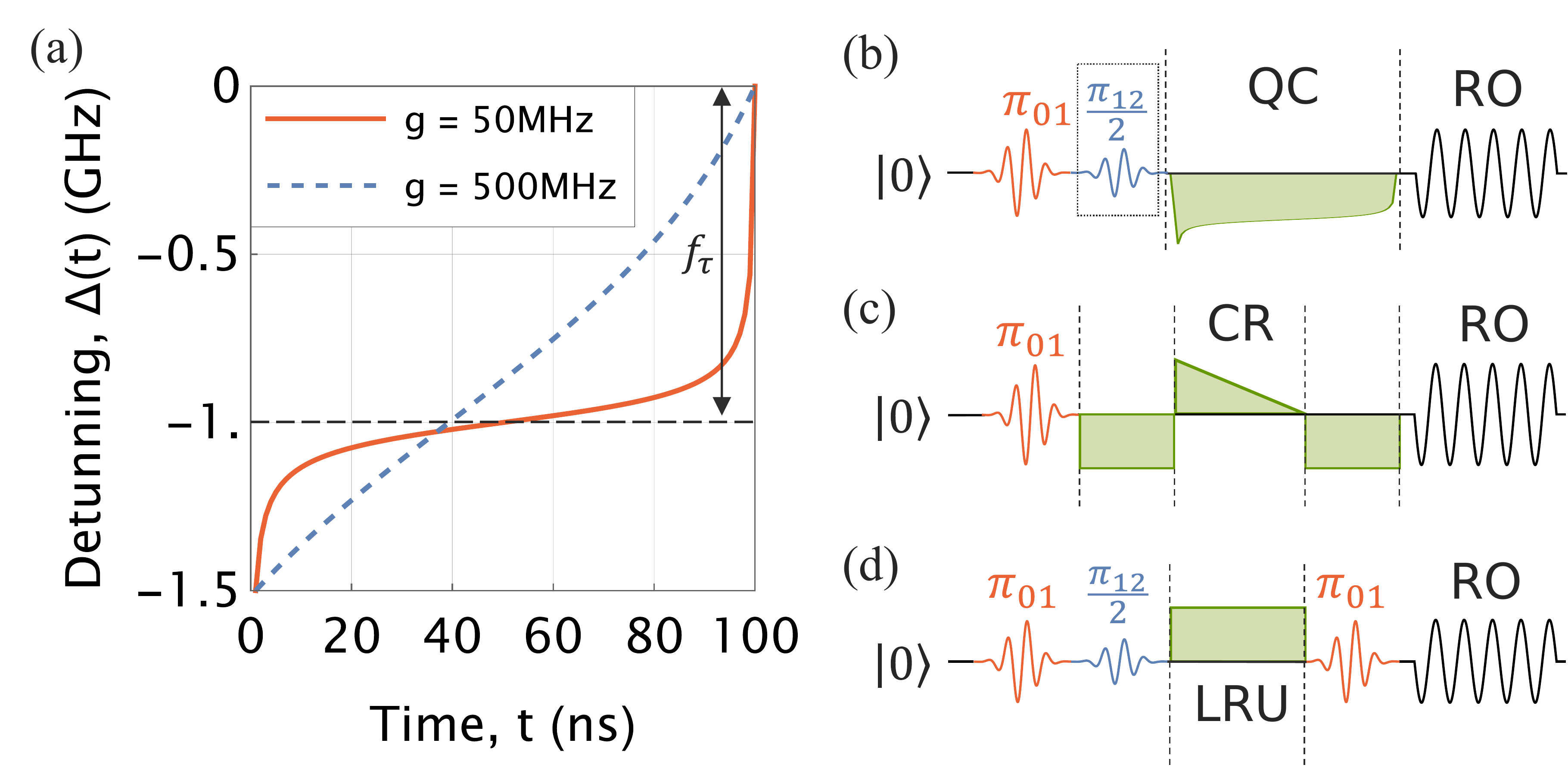}
 \caption{\textbf{Calibration sequences of the reset protocols with adiabatic pulses.} \textbf{(a)} The adiabatic pulse for the qubit-coupler (QC) reset SWAP is defined by four parameters: the pulse duration $\tau = 100~\unit{\nano\second}$, the coupler frequency offset, ${f_\tau} = -1~\unit{\giga\hertz}$, the QC coupling strength, $g$, and the QC detuning that defines the start of the adiabatic process, $f_{0} = -1.5~\unit{\giga\hertz}$. In particular, ${f_\tau}$ is essentially the amplitude of the flux pulse. The coupler is first pulsed to be below $\text{Q}_{0}$, and then crossing the $\text{Q}_{0}$ energy level adiabatically before returning to the idle position. The pulse schemes to calibrate the \textbf{(b)} qubit-coupler (QC) SWAP, \textbf{(c)} coupler-resonator (CR) SWAP and \textbf{(d)} leakage reduction unit (LRU) include $\pi_{01}$ (red) and $\pi_{12}/2$ (blue) pulses to prepare the qubit state, the flux pulses (filled areas in green) applied on the coupler, and the readout (RO) pulses on the resonator (black).}
    \label{fig:pulse_scheme}
\end{figure}

This condition implies that the changes in the interactions between the states $m$ and $n$ at time $t$ must be significantly smaller than the energy distance between the states. Note that this is a global condition and can be modified to an instantaneous condition. The instantaneous approximate adiabatic condition was previously used to accelerate Grover’s algorithm in adiabatic quantum computation by Roland and Cerf \cite{roland2002quantum, Albash2018}, which has been studied and experimentally verified in two-level systems \cite{stefanatos2020speeding, malossi2013quantum}. 

\begin{figure*}[ht!]
\centering
\adjustbox{lap={\width}{-1em}}{
\includegraphics[width=7.1in, keepaspectratio=true]{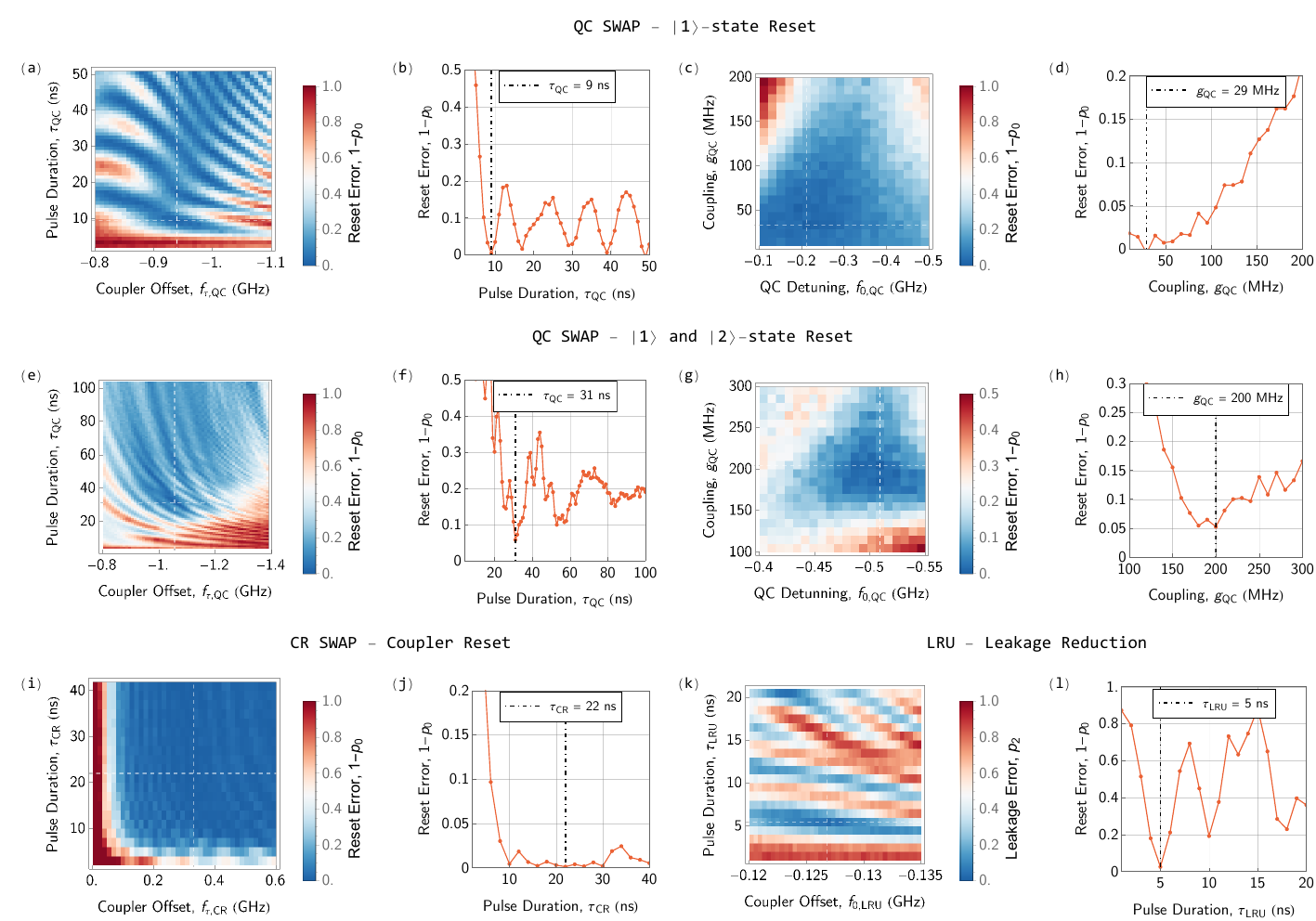}}

 \caption{\textbf{Calibration results of each step of the reset and leakage reduction protocol.} The measurement results of the \textbf{(a-h)} qubit-coupler (QC) SWAP, \textbf{(i,j)} coupler-readout resonator (CR) and \textbf{(k,l)} leakage reduction unit (LRU), respectively, are shown.
 The pulse sequence of each measurement is shown in  \figpanel{fig:pulse_scheme}{b-d}, respectively.
 For \textbf{(a-d)} the qubit is initialized in the $\estate$ state, while for \textbf{(e-h)} the initial state is $(\estate+\fstate)/\sqrt{2}$. For the 2D parameter sweeps, the colour represents the reset error, which takes all non-ground state populations into account, and the regions with lower reset errors are always indicated in blue. The line cut in \textbf{(b,d,f,h)} indicates the cross-section of the optimal parameters taken from the 2D sweep marked with a vertical white dashed line.}
    \label{fig:pulse_qc}
\end{figure*}

The exact analytical solution of the adiabatic pulse is derived in the Methods section, with an example shown and annotated in \figpanel{fig:pulse_scheme}{a}. The pulse can be defined by four parameters: $\tau$ is the pulse duration, $f_{0}$ is the coupler frequency offset from the idle point, $f_{\tau}$ is the qubit-coupler detuning at the start of the pulse, and $g$ is the qubit-coupler coupling strength that defines the slope of the center region of the pulse. In practice, the coupling parameter $g$ can be treated as an adjustable variable to be optimized for better reset performance. With increasing $g$, the pulse gradually transforms from a square pulse with an extra tail to a linear ramp pulse, so that we have the full range to tune the adiabaticity of the transition induced by this pulse. \par

\subsection*{Qubit-coupler SWAP gate}

The first step of our reset protocol is to transfer the $\estate$-state population of qubit $\text{Q}_{0}$ to coupler $\text{C}_{0}$, as shown in \figpanel{fig:pulse_scheme}{b}. We first bias the coupler $\text{C}_{0}$ to be at least 1~\unit{\giga\hertz} above $\text{Q}_{0}$ and below its resonator $\text{R}_{0}$, which would be the common idling point chosen for implementing a parametric two-qubit gate. We prepare $\text{Q}_{0}$ in $\estate$ (without the $\pi_{12}/2$-pulse at this stage) and then implement a QC SWAP operation between $\text{Q}_{0}$ and $\text{C}_{0}$ by applying the adiabatic flux pulse given by the solution of \eqref{eqn:global_adiabatic_condition} [see \eqref{eqn:detuning_full_solution} in Methods]. The pulse induces a population transfer between qubit and coupler, which is ideally adiabatic.\par
To tune-up the QC SWAP gate, we need to acquire the pulse parameters in two separate 2D sweeps, since four parameters are needed to define the pulse. For each sweep, we define the reset error $\epsilon_\text{reset}$ to be the residual non-ground-state population after the operation, $1 - p_{0}$, where $p_{0}$ is the measured $\gstate$-state population of $\text{Q}_{0}$. We use the reset error as the figure of merit for the calibration.\par
We first measure the reset error as a function of the pulse duration $\tau_{QC}$ and the coupler frequency detunning $f_{\tau}$. We set $f_{0}$ and $g$ to be 200~\unit{\mega\hertz} and 71~\unit{\mega\hertz}, respectively, given preliminary qubit and coupler spectroscopy results. The reset error as a function of $\tau_{QC}$ and $f_{\tau}$ is shown in \figpanel{fig:pulse_qc}{a}, and a chevron pattern illustrating population transfer from the qubit to the coupler $\estate$-state is observed. A line-cut of the first minima in reset error is displayed in \figpanel{fig:pulse_qc}{b} to demonstrate the time evolution of the population. We then fix $\tau_{QC}$ and $f_{\tau}$ to be the values that produce the lowest reset error, and sweep both $g$ and $f_{0}$ to refine the pulse shape, as shown in \figpanel{fig:pulse_qc}{c-d}. A large parameter space, with $g$ being below 50~\unit{\mega\hertz} and $f_{0}$ being around -200~\unit{\mega\hertz}, can be identified where the reset error approaches the readout limit. These initial results suggest that the fidelity of this swap operation can be above 99\%, limited by drifting qubit parameters and coupler flux bias currents. We achieve depopulation of the $\estate$ state with a 9~\unit{\nano\second} QC SWAP gate between the ancilla qubit and the coupler. The data qubit, $\text{Q}_{1}$,  is only negligibly affected by the flux pulse, since it is 642~\unit{\mega\hertz} lower in frequency than the ancilla qubit. Therefore, the reset of the ancilla qubit can be performed independently of the state of the data qubit; more details can be found in Supplementary Note II.\par
To show how we can reset both the $\estate$-state and $\fstate$-state population of the ancilla qubit simultaneously, we prepare the ancilla in a superposition state $(\estate+\fstate)/\sqrt{2}$ with an additional $\pi_{12}/2$-pulse [see \figpanel{fig:pulse_scheme}{b}]
so that the effect on both states is visible with a single measurement. The coupler interacts with multiple qubit levels during the frequency tuning see \figref{fig:energy_simulation} in the Methods section). The results are shown in \figpanel{fig:pulse_qc}{e-h}. There is a relatively small parameter space for a fast reset on the order of 31~\unit{\nano\second} due to interaction between both qubit $\estate$- and $\fstate$-state population with the coupler, with around $5\%$ of the population remaining at the $\estate$ state while the rest is in the $\gstate$ state. This is due to the proximity of the data qubit $\text{Q}_{1}$ to $\text{Q}_{0}$ in frequency, limiting the furthest extent of the adiabatic pulse and thus the adiabadicity of the pulse. Therefore, we are able to reset both the $\estate$ and the $\fstate$ states with the same pulse parameters, although the reset of the $\fstate$ state is only partially complete with a single pulse. \par
To improve the reset fidelity, we repeat the same reset scheme with another coupler. In practice, the implementation is realized with coupler $\text{C}_{0}$ and coupler $\text{C}_{2}$, both of which are coupled to qubit $\text{Q}_{0}$. Assuming that the $\estate$-state reset is ideal, the reset success probability is $1-2\cdot0.05=90\%$ for a pure $\fstate$ state. From this, we can estimate that two successive reset operations with similar success probability can achieve a total fidelity of $1-0.1^{2}=99\%$, albeit at the cost of double the time.\par

\subsection*{Coupler-resonator SWAP gate}

The next step of the reset protocol is to remove the excitation in the coupler by resonant interaction with the qubit readout resonator, as shown in \figref{fig:energy}. Due to the direct capacitive coupling between the coupler and the qubit readout resonator and indirect coupling mediated by the qubit, it is possible to implement this scheme without changing the architecture of the device.\par


To calibrate the coupler-resonator SWAP, we implement the pulse sequence shown in \figpanel{fig:pulse_scheme}{c}. We first prepare the ancilla qubit in the $\estate$ state and then apply the reset flux pulse, starting with a QC SWAP to populate the coupler. A diabatic square pulse is preferable here during the calibration since it is easier to observe the oscillation of the exciation between the qubit and the coupler. Afterwards, we implement a linear-ramp pulse to perform the CR SWAP to induce adiabatic transfer, due to the simpler nature of the frequency landscape of the resonators. The amplitude of the CR SWAP flux pulse has the opposite sign of the QC SWAP pulse in order to interact with the resonator levels that are higher in frequency. We add an extra QC SWAP pulse after the CR SWAP pulse to move the residual population in the coupler back to the ancilla qubit so that its population can be read out.\par

The coupler is reset to its ground state $\ket{\text{0}_{c}}$  after 20~\unit{\nano\second}, as shown in \figpanel{fig:pulse_qc}{i-j}, with reset error reaching below the readout floor at $10^{-4}$ for a relatively large parameter space. We choose a pulse duration of 22~\unit{\nano\second} for robustness against fluctuations over time. Note that the CR SWAP can be applied to all couplers simultaneously to reduce sequence time.\par


\subsection*{Leakage reduction unit with diabatic pulse}

Leakage in the data qubits is a major error source in any error-correction scheme. Therefore, a protocol to specifically target the $\fstate$ state of data qubits without disrupting the $\estate$-state population is necessary for successful error correction. A LRU on the data qubit $\text{Q}_{1}$ can be implemented by moving the coupler $\text{C}_{1}$ to where the $\ket{\text{1}_{\text{1}}\text{1}_{\text{c}}}$ and $\ket{\text{2}_{\text{1}}\text{0}_{\text{c}}}$ states of the $\text{Q}_{1}$-$\text{C}_{1}$ system are on resonance, as shown in \figref{fig:energy}. The coupler $\text{C}_{1}$ that implements the LRU  is parked below the data qubit, to avoid disruption of its $\estate$-state population.\par

The specific pulse scheme is shown in \figpanel{fig:pulse_scheme}{d}. We prepare $\text{Q}_{1}$ in the $(\ket{\text{1}}+\ket{\text{2}})/\sqrt{2}$ superposition state, and apply the flux pulse to move the coupler $\text{C}_{1}$ such that the $\ket{\text{2}_{\text{1}}\text{0}_{\text{c}}}$ state is on resonance with the $\ket{\text{1}_{\text{1}}\text{1}_{\text{c}}}$ state. The LRU pulse shape is chosen to be a simple square pulse such that the interaction is fast and diabatic. This minimizes the effect on the qubit $\estate$-state population. An extra $\pi_{01}$ pulse is added after the LRU, since the readout fidelity is higher when measuring the $\gstate$-state population. The results in \figpanel{fig:pulse_qc}{k-l} show that it takes 5~\unit{\nano\second} to complete the population swapping from the $\ket{\text{2}_{\text{1}}\text{0}_{\text{c}}}$ to the $\ket{\text{1}_{\text{1}}\text{1}_{\text{c}}}$ state. \par
Furthermore, the LRU can be implemented simultaneously with the reset pulse on the ancilla qubit, since it uses a different coupler, $\text{C}_{1}$. Similarly to the reset protocol, after interacting with the qubits, all couplers are made to adiabatically interact with the resonators simultaneously to transfer the couplers' population.\par

\subsection*{Single-shot verification}

\begin{figure}[t!]
\centering
  \includegraphics[width=1.6in, keepaspectratio=true]{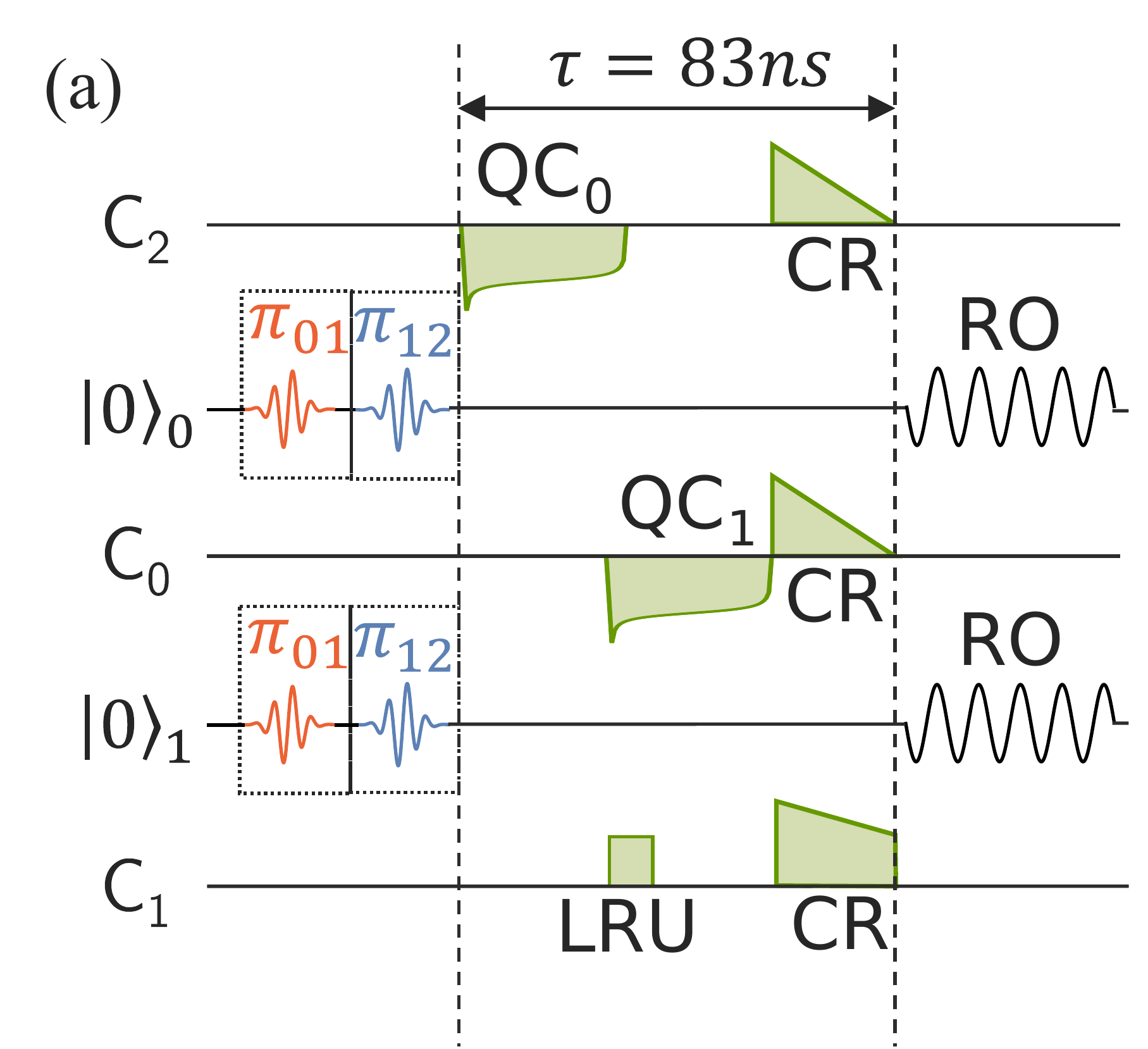}
  \includegraphics[width=1.7in, keepaspectratio=true]{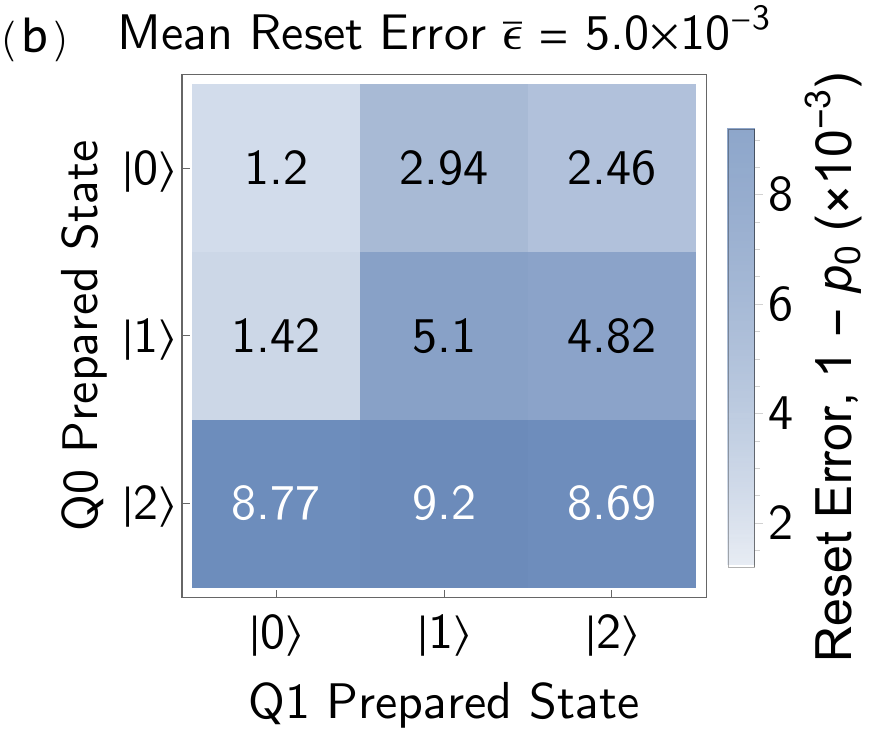}
  \includegraphics[width=3.3in, keepaspectratio=true]{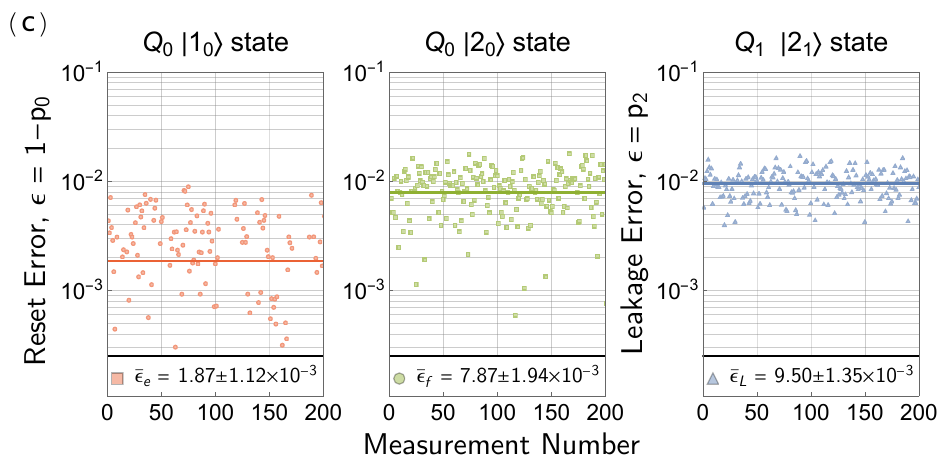}
 \caption {\textbf{Pulse sequence and performance verification with single-shot readout.} \textbf{(a)} The pulse sequences include state preparation of two qubits, flux pulses applied on three couplers for reset and leakage reduction, and the readout of qubit population. The total duration of the flux pulse sequence is 83~\unit{\nano\second}. All 9 combinations of two-qubit initial states in the first and second excitation manifold are prepared and measured. 
 \textbf{(b)} Single-shot readout results of resetting all 9 two-qubit states. The mean reset error is calculated from the average of all combinations. \textbf{(c)} Long-term single-shot readout results for the 3 out of 9 combinations where $\text{Q}_{0}$ is prepared in the $\estate$- or $\fstate$-state, or $\text{Q}_{1}$ is prepared in the $\fstate$-state respectively. The measurement is repeated 200 times to test the stability of the protocol.}
\label{fig:ssro}
\end{figure}

To fully characterize the reset and LRU performance, we prepare the data and ancilla qubits in all nine combinations of two-qubit states in the first and second excitation manifolds, i.e., $\ket{\text{0}_{\text{0}}\text{0}_{\text{1}}}$, $\ket{\text{0}_{\text{0}}\text{1}_{\text{1}}}$, $\ket{\text{0}_{\text{0}}\text{2}_{\text{1}}}$, etc. Then we execute the full reset protocol, including $\estate$- and $\fstate$-state QC SWAP, CR SWAP, and LRU with the two qubits ($\text{Q}_{0}$ and $\text{Q}_{1}$) and the three couplers ($\text{C}_{0}$, $\text{C}_{1}$, and $\text{C}_{2}$), as shown in \figpanel{fig:ssro}{a}. With the measurement protocol established, we employ the CMA-ES algorithm to optimize the pulse parameters $g$, $f_{\tau}$, and $f_{0}$ to achieve a lower reset error \cite{Hansen2005}.\par

Time-wise, resetting only the $\estate$ state of the ancilla qubit requires only 9~\unit{\nano\second}. However, to reset both the $\estate$ and $\fstate$ states, we need to apply flux pulses on couplers $\text{C}_{0}$ and $\text{C}_{2}$ sequentially; these pulses are 31~\unit{\nano\second} and 30~\unit{\nano\second} long, respectively. The CR SWAP takes 22~\unit{\nano\second} while the leakage reduction unit needs a 5-ns-long pulse, which can start at the same time as the QC SWAP pulses. Therefore, the total time needed to implement both reset and LRU is 83~\unit{\nano\second}.\par

Each of the nine states is prepared and measured for 10,000 shots with single-shot readout. The results for each measurement are illustrated as a matrix in \figpanel{fig:ssro}{b}. To understand the long-term stability and error margin, the results are gathered from 200 repeated measurements as illustrated in \figpanel{fig:ssro}{c}. We average over the three out of nine combinations of the initial state where $\text{Q}_{0}$ is prepared in the $\estate$ or $\fstate$ state, or  $\text{Q}_{1}$ is prepared in the $\fstate$ state, to trace out the effect of the different input states. On average, the adiabatic reset protocol achieves a reset error of $1.87\times10^{-3}$ ($7.87\times10^{-3}$) for the $\estate$ ($\fstate$) state. Meanwhile, the leakage reduction reaches an error of $9.50\times10^{-3}$. Over time, all three components are robust against random fluctuations during the span of at least a few hours.\par

\section*{DISCUSSION}

In summary, we have implemented a fast, high-fidelity, and unconditional qubit reset protocol on fixed-frequency qubits with a tunable coupler. Using adiabatic pulses, our reset protocol achieves a reset error of $(1.87\pm1.12)\times10^{-3}$ for the $\estate$ state within 9~\unit{\nano\second}, and $(7.87\pm1.94)\times10^{-3}$ for the $\fstate$ state in 61~\unit{\nano\second}. We also perform leakage reduction, on the qubit in 5~\unit{\nano\second} with a remaining leakage error of $(9.50\pm1.35)\times10^{-3}$. Afterwards, the population in the coupler can be transferred to the readout resonator in 22~\unit{\nano\second}. In total, the combination of qubit reset, leakage reduction and coupler reset takes only 83~\unit{\nano\second} to complete. The reset error we achieved is below the suggested threshold for quantum error correction \cite{geher2024}, which is between $10^{-2}$ to $10^{-2.5}$, together with the additional benefit of removing leakage in both aniclla and data qubits. The reset pulses are straightforward to tune up, with at most four parameters completely defining the entire pulse shape. Moreover, the reset of the ancilla qubits and the leakage reduction unit of the data qubits can run simultaneously to achieve maximal efficiency due to the usage of all available coupler elements. \par

The main limitation on the fidelity of the reset protocol is found to be the frequency separation of the qubit pair. With only 642~\unit{\mega\hertz} of spacing between the two qubits, the adiabatic pulses have a limited frequency range to evolve back to the initial state of the coupler in the dispersive regime, which leads to an incomplete adiabatic transition. Another limiting factor for the LRU is the small anharmonicity of the couplers on the current device, which affects the undesirabled interaction with the $\estate$ state of the data qubits. Further design iterations and development in pulse-shaping techniques to alleviate these undesirable effects are currently under investigation with more theoretical simulations and better parameter-optimization algorithms. \par

To see how to scale up the reset and LRU protocol in the surface code implementation based on a 2D square grid, we can start by estimating the ratio between the number of qubits and couplers. For a code distance $d$, there will be $d^2$ data qubits, $d^2-1$ ancilla qubits, and $c_{d} = 4d(d-1)$ couplers. The reset and LRU protocol will need $c_{r} = 3\cdot d^2 -1$ couplers to fully implement the scheme. For $d>3$, $c_{d}$ is always greater than $c_{r}$, thus guaranteeing the implementation of our protocol without the need for additional elements. Moreover, we have recently demonstrated \cite{Kosen2024} that our processor has a low level of crosstalk, thus, enabling us to further scale up our design. In conclusion, we have demonstrated that the architecture with fixed-frequency qubits and tunable couplers is compatible with quantum error-correction schemes and subsequent fault-tolerant quantum computing.\par






\section*{METHODS}
\subsection*{System Hamiltonian}

We first consider a system consisting of a fixed-frequency transmon qubit, a tunable transmon that acts as the coupler element, and a leaky resonator forming the dissipator stage. Invoking the rotating-wave approximation (RWA), we model the qubit and the coupler as Kerr oscillators with anharmonicities $\alpha_{q}$ and $\alpha_{c}$, respectively, and we write the system Hamiltonian as\par
\begin{equation}
\label{eqn_RateEfPulse}
\begin{split}
H_\text{RWA} = & \: \omega_{r}a^{\dagger}a+\omega_{q}b^{\dagger}b+\frac{\alpha_{q}}{2}b^{\dagger2}b^{2}+\omega_{c}c^{\dagger}c+\frac{\alpha_{c}}{2}c^{\dagger2}c^{2}\\
& \: +g_{qr}\left(a^{\dagger}b+b^{\dagger}a\right)+g_{qc}\left(b^{\dagger}c+c^{\dagger}b\right),\\
\end{split}
\end{equation}
where $a$ ($a^{\dagger}$), $b$ ($b^{\dagger}$), $c$ ($c^{\dagger}$) are the annihilation (creation) operators for the resonator, qubit, and coupler, labeled by the subscripts $r$, $q$, and $c$, respectively. $g_{ij}$ denotes the coupling strengths between the systems $i$ and $j$. For the theoretical study, we assume that the coupler is not directly connected to the resonator. Therefore, the exchange of quanta between the coupler and resonator primarily occur near their resonance, via second-order interactions that can be studied using the well-known Schrieffer-Wolff perturbative expansion \cite{Bravyi2011}.\par

Note that $H_\text{RWA}$ is number-conserving as all the terms contain an equal number of raising and lowering ladder operators. The Hamiltonian decouples into different blocks labelled by the total number of excitations $N$, and each block can be studied separately. The dynamics of a qubit starting in its first-excited state can be studied within the first-excitation subspace:\par

\begin{equation}
\label{eqn_1ExHamiltonian}
H_{1}= \begin{bmatrix}
 \omega_{r} &g_{rq}  & 0 \\
g_{rq} & \omega_{q} & g_{qc} \\
0 & g_{qc} & \omega_{c}
\end{bmatrix}.
\end{equation}

\subsection*{Diabatic SWAP interactions}

To understand the dynamics of the reset, we start by examining the case of diabatic swap between the qubit and the coupler, which contributes to reset the $\estate$-state population of the qubit. We employ here both analytical and numerical tools to understand the reset scheme when the coupler frequency is brought into resonances on time scales shorter than the time scale $1/g$. The first step of an adiabatic swap is to bring the coupler and the qubit into resonance. We achieve this by tuning the coupler frequency. During the waiting time at resonance, the qubit exchanges excitations with the coupler, thus completing the first QC SWAP.\par

At the QC resonance, the resonator is far detuned from the qubit ($\omega_{r}-\omega_{q} \gg g_{rq}$) such that we neglect its contribution to the dynamics in the first swap interaction. This makes the analysis even simpler as it reduces the three-level system in \eqref{eqn_1ExHamiltonian} to an effective two-level system (TLS) formed by the qubit and the coupler near resonance. Thus, we expect Rabi oscillations between the qubit and coupler.\par

We start the protocol with the system initialized in $\ket{\text{1}_{\text{q}}\text{0}_{\text{c}}\text{0}_{\text{r}}}$ where states are labeled in the order $\ket{\text{qcr}}$. The effective TLS Hamiltonian in the space of states $\ket{\text{1}_{\text{q}}\text{0}_{\text{c}}\text{0}_{\text{r}}}$ and $\ket{\text{0}_{\text{q}}\text{1}_{\text{c}}\text{0}_{\text{r}}}$ interacting in the first QC SWAP is\par

\begin{equation}
\label{eqn_RateEfPulse}
H_{qc}=\begin{bmatrix}
-\Delta_{qc}/2 & g_{qc} \\
 g_{qc}&\Delta_{qc}/2 
\end{bmatrix},
\end{equation}
with $\Delta_{qc} = \omega_{q}-\omega_{c}$.

During the waiting time $\tau$ of the first swap, $\Delta_{qc} = 0$. Therefore, the state at time $t$ is 
\begin{equation}
\label{eqn_psi_t}
\left| \Psi_{t} \right\rangle = \cos (gt)\left| 1_{q}0_{c}0_{r}\right\rangle+i\sin (gt)\left| 0_{q}1_{c}0_{r} \right\rangle,
\end{equation}
with the qubit population entirely moving to the coupler at time $t = \pi / 2g$. The interaction is ideally an iSWAP gate. However, since we only consider the population transfer during reset, and we choose to simplify the notation to be SWAP. After the first swap, the coupler is in the excited state, meaning that the coupler needs to move away from resonance near $t = \pi / 2g$. This marks the end of the first swap. The coupler frequency is then ramped up to the resonator frequency to initiate the second step, to enable the interaction between the resonator and the coupler. At resonance, the coupler and the resonator undergo Rabi oscillations. The lossy resonator leaks out almost completely in a time $\approx 3/\kappa$, thus emptying the resonator and completing the full reset operation.\par

The TLS Hamiltonian for the CR SWAP is

\begin{equation}
\label{eqn_RateEfPulse}
H_{cr}=\begin{bmatrix}
-\Delta_{cr}/2 & g_{cr} \\
 g_{cr}&\Delta_{cr}/2 
\end{bmatrix},
\end{equation}
with $\Delta_{cr} = \omega_{c}-\omega_{r}$. The resonator has a loss rate $\kappa$, which makes the evolution non-unitary and necessary to treat within the master-equation formalism. If we assume that the coupler and the resonator are not simultaneously excited, the master equation implies that the non-Hermitian Hamiltonian
\begin{equation}
\label{eqn_RateEfPulse}
H_\text{eff}=H-\mathrm{i}\frac{\kappa}{2} \mathcal{L}^{\dagger}\mathcal{L}
\end{equation}
equivalently accounts for the dissipation. Solving its effective Schrödinger yields an analytical expression for the non-unitary evolution of the resonator and coupler states. Here, we consider photon loss from the cavity as the main source of dissipation. Therefore, the Lindblad operator $\mathcal{L}$ annihilates a resonator excitation, i.e., $\mathcal{L}=a$.
The effective non-Hermitian Hamiltonian governing the second swap can be written as 

\begin{equation}
\label{eqn_hamiltonian_cr}
H_{cr}=\begin{bmatrix}
-\Delta_{cr}/2 & g_{cr} \\
 g_{cr}&\Delta_{cr}/2-\mathrm{i} \frac{\kappa}{2}
\end{bmatrix}.
\end{equation}

Solving the Schrödinger equation for the Hamiltonian in \eqref{eqn_hamiltonian_cr}, we find the decay of the excited coupler wavefunction $\psi_{c}$:\par

\begin{equation}
\begin{cases} 
\frac{e^{-\frac{\kappa t}{4}}}{2g_{cr}} \left( \kappa \sinh\left( \frac{\sqrt{|\alpha|} t}{4} \right) + \sqrt{|\alpha|} \cosh\left( \frac{\sqrt{|\alpha|} t}{4} \right) \right),&\alpha > 0 \\
\frac{e^{-\frac{\kappa t}{4}}}{2g_{cr}} \left( \kappa \sin\left( \frac{\sqrt{|\alpha|} t}{4} \right) + \sqrt{\alpha} \cos\left( \frac{\sqrt{|\alpha|} t}{4} \right) \right),&\alpha < 0 \\
\frac{e^{-\frac{\kappa t}{4}}}{g_{cr}} \left( \frac{\kappa t}{4} + 1 \right),&\alpha = 0
\end{cases}
\label{eqn_coupler_wavefunction}
\end{equation}
with $\alpha = \kappa^2 - 4g_{cr}^2$. This provides important insights into the overall reset rates. One might expect that the photon decay rate and, thus, the reset speed increases with $\kappa$. As seen from \eqref{eqn_coupler_wavefunction}, the reset can be divided into three regimes, under-damped $(\alpha < 0)$, over-damped $(\alpha > 0)$, and critically damped $(\alpha = 0)$. In the over-damped regime, populations decay without oscillations. The decay rate is less than $\kappa/2$ as the terms inside the parenthes, contribute to population growth and slow down the decay. In the under-damped regime, the decay is oscillatory, limiting the overall speed of population decay. The reset rate increases with $\kappa$ until it hits the critical point $\kappa/g_{cr} = 2$. At the critical damping point, the population decays without oscillations. This is where we expect to get the fastest decay at a rate $\kappa/2$. The ratio $\kappa/g_{cr} = 2$ suggests the optimal point of operation in terms of the overall reset speed.\par
From Eqs.~(\ref{eqn_psi_t}) and (\ref{eqn_coupler_wavefunction}), we can derive the total reset time for the fastest case of diabatic reset when it is operated in the regime of  the critical damping. At resonance, the Rabi frequency for CR exchange is $\Omega_{R} = 2g_{cr}$. Similarly, the Rabi frequency of the QC interaction is $\Omega_{Q} = 2g_{qc}$. Using the above relation, we define the total reset time to be:

\begin{equation}
\label{eqn_RateEfPulse}
T_{reset} = \Omega_{R}^{-1}+\Omega_{Q}^{-1}+\kappa^{-1},
\end{equation}
which can be used to estimate the reset speed for future device designs. To enable a total reset time on the order of 100 of ns, a set of typical design parameters can be $\{\Omega_{R}/2\pi,\Omega_{Q}/2\pi,\kappa/2\pi\} = \{10,60,10\}$~\unit{\mega\hertz}, which is feasible with a flip-chip architecture and Purcell-filter designs for fast readout. 

\subsection*{Simulation of two-qubit-coupler subsystem}

\begin{figure}[t!]
\centering
 \includegraphics[width=3.4in, keepaspectratio=true]{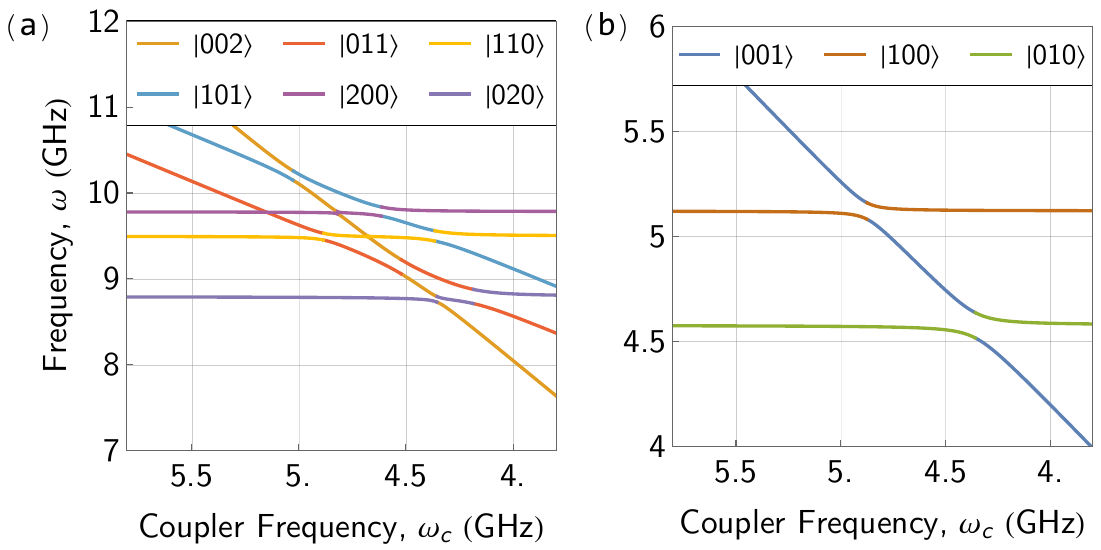}
 \caption{\textbf{Simulated energy spectrum of the $\text{Q}_{0}$-$\text{Q}_{1}$-$\text{C}_{0}$ system and QC SWAP with a square flux pulse.} \textbf{(a)}  Energy spectrum that involves two-excitation states such as $\ket{002}$, $\ket{011}$, $\ket{110}$, $\ket{101}$, $\ket{200}$, and $\ket{020}$ as a function of the coupler frequency. These states represent different combinations of excitations distributed among the two qubits and the coupler. \textbf{(b)} Energy spectrum of the single-excitation states $\ket{001}$, $\ket{100}$, and $\ket{010}$, which involve a single excitation localized in either $\text{Q}_{0}$, $\text{Q}_{1}$, or $\text{C}_{0}$.}
    \label{fig:energy_simulation}
\end{figure}

We simulate the effect of the coupler frequency-tuning to understand the system behaviour in the regime when the coupler is interacting with both qubits \cite{Fors2024}. The simulated energy spectrum of the $\text{Q}_{0}$-$\text{Q}_{1}$-$\text{C}_{0}$ system is shown in \figref{fig:energy_simulation}, consisting of the two fixed-frequency transmon qubits, $\text{Q}_{0}$ and $\text{Q}_{1}$, coupled via the tunable coupler $\text{C}_{0}$. In this architecture, the coupler frequency ($\omega_c$) can be adjusted to control the interactions between the qubits, enabling dynamical control over the energy levels of the system. As the coupler frequency is tuned, the energy levels shift, displaying several avoided level crossings where the states interact strongly, highlighting the tunable coupling mechanism.\par

The avoided crossings between the states $\ket{200}$, $\ket{101}$, and $\ket{002}$ in \figpanel{fig:energy_simulation}{a}, as well as $\ket{100}$ and $\ket{001}$ in \figpanel{fig:energy_simulation}{b}, are of particular interest as they provide a pathway for adiabatic state transfer between qubit $\text{Q}_{0}$ and the coupler $\text{C}_{0}$. By slowly varying the coupler frequency, the system can transition between these states without occupying intermediate levels, facilitating a controlled swap of a excitation from qubit $\text{Q}_{0}$ to the coupler. This mechanism is crucial for our approach in unconditional reset of the qubit $\text{Q}_{0}$.\par

\begin{figure}[t!]
\centering
 \includegraphics[width=3.4in, keepaspectratio=true]{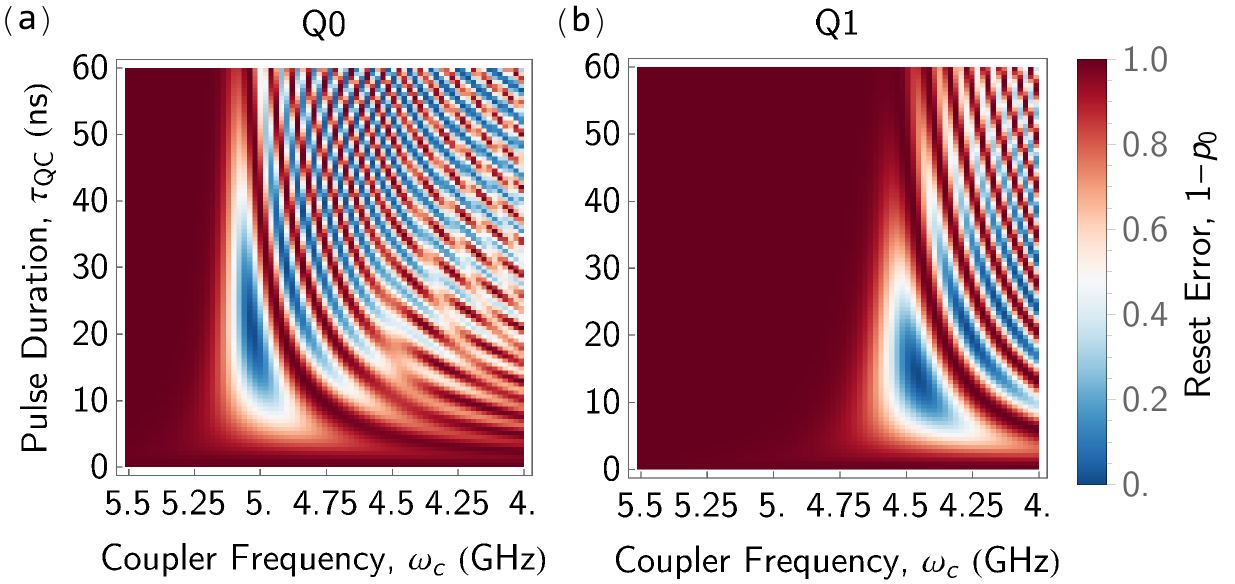}
 \caption{\textbf{QC SWAP with a diabatic flux pulse.} Reset error for the qubit \textbf{(a)} 
 $\text{Q}_{0}$ and \textbf{(b)} $\text{Q}_{1}$. The color scale represents the reset error, defined as $1 - p_0$, where $p_0$ is the probability of the qubit being in the ground state after the reset protocol. The heatmaps illustrate how the state-transfer dynamics and subsequent reset errors depend on the interaction between the qubits and the tunable coupler. Darker regions (red) indicate higher reset errors, while lighter regions (blue) signify more efficient resets with lower error rates.}
    \label{fig:reset_simulation}
\end{figure}

We then simulate reset errors of qubits $\text{Q}_{0}$ and $\text{Q}_{1}$ as a function of the coupler frequency ($\omega_c$) and pulse duration ($\tau_{QC}$) for a square flux pulse. The result is shown in \figref{fig:reset_simulation}. The simulation result shows that $\text{Q}_{1}$ is not affected by the reset pulse given that one chooses $\omega_{c}$ and $\tau_{QC}$ to best reset $\text{Q}_{0}$.\par

However, due to the anharmonicity of the transmon, the higher energy transitions have a multitude resonant interactions with the coupler states as shown in \figpanel{fig:energy_simulation}{a}. It becomes difficult to reset both the $\estate$- and $\fstate$-state population of the qubit with a single diabatic pulse. Therefore, we need to examine the feasibility to reset multiple qubit states with adiabatic transfer instead.

\subsection*{Adiabatic Landau-Zener-Stückelberg transitions}
To understand the adiabatic component of the reset dynamics, we consider the Landau-Zener-Stückelberg (LZS) problem \cite{Stuckelberg1932,Landau1932, IVAKHNENKO20231}. We will first describe the LZS problem and then map our system to it. Consider a TLS described by the following Hamiltonian, which varies linearly in time:


\begin{equation}
\label{eqn_RateEfPulse}
\begin{split}
H_\text{LZS}=\begin{bmatrix}
\alpha t/2 & 0 \\
 0&-\alpha t/2 
\end{bmatrix} &,~
E_{\pm} = \pm \alpha t/2 ,\\
\Psi_{+}=\begin{bmatrix}
1 \\0
\end{bmatrix}&,~
\Psi_{-}=\begin{bmatrix}
0 \\1
\end{bmatrix}.
\end{split}
\end{equation}

Since the Hamiltonian is diagonal, solving the time evolution of the instantaneous eigenstates is
straightforward ($\hbar=1$):
\begin{equation}
\label{eqn_RateEfPulse}
\left| \Psi_{\pm}(t) \right\rangle = e^{\pm i \alpha t^2/2}\ket{\Psi_{\pm}}.
\end{equation}
After adding a coupling strength along the off-diagonals of the Hamiltonian

\begin{equation}
\label{eqn_RateEfPulse}
H_\text{LZS}=\begin{bmatrix}
\alpha t/2 & g \\
 g&-\alpha t/2 
\end{bmatrix} ,~
E_{\pm} = \pm \sqrt{\left( \frac{\alpha t}{2}^2 \right) + g^2},
\end{equation}
the before eigenstates $\left| \Psi_\pm \right\rangle$ are no longer time-independent, which leads to $\Psi_{+} \to \Psi_{-}$ and $\Psi_{+} \to \Psi_{-}$.

\begin{figure}[t!]
\centering
 \includegraphics[width=3.3in, keepaspectratio=true]{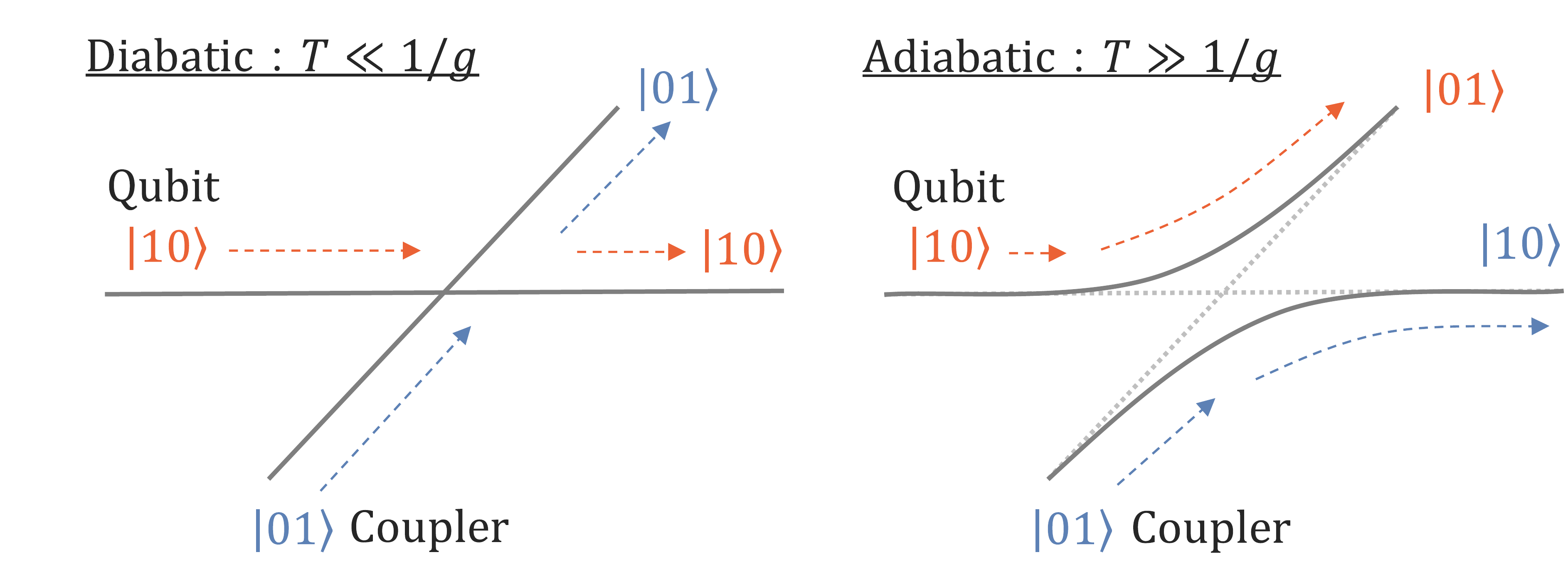}

 \caption{\textbf{Diabatic and adiabatic transitions.} The qubit and coupler energy levels are coupled with coupling strength $g$. The qubit level is stationary while the coupler is transiting across for a duration $T$. There are two limiting scenarios in this case: 1) in fast diabatic transition (left), $gT<1$, the population stays on its original trajectory, 2) in slow adiabatic transition (right), $gT\gg1$, the population is swapped to the other energy level during the interaction. }
    \label{fig:transition}
\end{figure}

The minimum energy gap between the eigenstates occurs at $t = 0$, $\delta_\text{min}=2g$. This sets the time scale for
transitions, and two limiting scenarios arise, as shown in \figref{fig:transition}. If the transition duration is on a timescale of $T \gg 1/g$, the process is said to be adiabatic. This is the origin of the ramp-up time set to $T < 1/g$ in the diabatic swap interactions. It can be shown that the
probability of a transition, a non-adiabatic effect, is

\begin{equation}
\label{eqn_RateEfPulse}
P_\text{trans} = e^{-\frac{2\pi g^2}{\hbar \alpha}}.
\end{equation}

This clearly indicates that the evolution becomes more adiabatic for larger values of couplings, which
can be attributed to the increase in the energy gap at the avoided level crossing.

\subsection*{Adiabatic pulse shape}

Following the Roland and Cerf protocol to speed up the adiabatic transitions \cite{roland2002quantum, Albash2018}, the precise adiabatic pulse shape can be derived by assuming that the adiabatic condition in \eqref{eqn:global_adiabatic_condition} is satisfied for every inﬁnitesimal time interval from $t$ to $t+dt$, arriving at
\begin{equation}
\left|\left\langle\Psi_n(t)\left|\frac{\partial H(t)}{\partial t}\right| \Psi_m(t)\right\rangle \right| \ll |E_n(t) - E_m(t)|^2.
    \label{eqn:local_adiabatic_condition}
\end{equation}

For a two-level system (TLS) with coupling strength $g$ that is described by the Hamiltonian
\begin{equation}
H_{\rm TLS}(t)= \hbar \begin{bmatrix}
\Delta(t)/2 & g \\
 g & -\Delta(t) / 2 
\end{bmatrix},
\label{eqn:tls_hamiltonian}
\end{equation}
the adiabatic condition, defined by $\eqref{eqn:local_adiabatic_condition}$, is:
\begin{equation}
\left|\frac{\partial \Delta(t)}{\partial t}\right| \ll \frac{(\Delta^2(t) + 4g^2)^{\frac{3}{2}}}{g} . 
\label{eqn:detuning_adiabatic_condition_inq}
\end{equation}

We find the qubit-coupler detuning $\Delta(t)$ for adiabatic evolution. The solution to the differential equation obtained by multiplying a scaling prefactor $\beta$ to the right-hand side of \eqref{eqn:detuning_adiabatic_condition_inq} is:
\begin{equation}
    \left|\frac{\partial \Delta(t)}{\partial t}\right| = \beta\frac{(\Delta^2(t) + 4g^2)^{\frac{3}{2}}}{g}.
\label{eqn:detuning_adiabatic_condition_a}
\end{equation}

Therefore, given the boundary conditions for a pulse duration of $\tau$, which are denoted as $\Delta(0)=f_0$ and $\Delta(\tau)=f_\tau$, one can find the solution to \eqref{eqn:detuning_adiabatic_condition_a} for the instantaneous adiabatic evolution to be
\begin{align}
    \Delta(t) &= -\frac{8g(\beta g \cdot t + \delta)}{\sqrt{1 - 16(\beta g \cdot t)^2 - 32\beta \delta g \cdot t  - 16 \delta^2}},
    \label{eqn:detuning_full_solution}
\end{align}
with
\begin{align}
    \beta &= \frac{-4\delta(f_\tau^2 + 4g^2) - f_\tau\sqrt{f_\tau^2 + 4g^2})}{4g\tau(f_\tau^2 + 4g^2)},\\
    \delta &= -\frac{f_0}{\sqrt{16f_0^2 + 64g^2}}.
    \label{eqn:detuning_full_solution_a_c}
\end{align}
Trajectories similar to \eqref{eqn:detuning_full_solution} based on the Roland and Cerf protocol can also be found in Refs. \cite{roland2002quantum, bason2012high, malossi2013quantum, stefanatos2019resonant, petiziol2019accelerating}.

\clearpage

\section*{Data availability}

All relevant data and figures supporting the main conclusions of the document are available on Zenodo. Please refer to Liangyu Chen at liangyuc@chalmers.se if needed.\par

\section*{Code availability}

All relevant code supporting the document is available online via \href{https://github.com/tergite/tergite-autocalibration}{GitHub}. Please refer to Liangyu Chen at liangyuc@chalmers.se if needed.\par

\section*{Acknowledgements}

We would like to thank Per Delsing, Christopher W. Warren, Christian J. Križan, Anuj Aggarwal, Janka Biznárová, Tangyou Huang, Guangze Chen and Huizhong Cao for valuable discussions and insights. The fabrication of our quantum processor was performed in part at Myfab Chalmers and flip-chip integration was done at VTT. The device simulations were enabled by resources provided by the Swedish National Infrastructure for Computing (SNIC) at National Supercomputer Centre (NSC) partially funded by the Swedish Research Council through grant agreement no. 2018-05973. This research was financially supported by the Knut and Alice Wallenberg Foundation through the Wallenberg Center for Quantum Technology (WACQT), the Swedish Research Council, and by the EU Flagship on Quantum Technology HORIZON-CL4-2022-QUANTUM-01-SGA project 101113946 OpenSuperQPlus100. A.F.K. also acknowledges support from the Swedish Foundation for Strategic Research (grant numbers FFL21-0279 and FUS21-0063).\par

\section*{Author Contributions}

L.C. performed the measurements and analysis of the results; S.K., H.L., D.S. and A.O. designed and simulated the device; L.C., B.L., and S.P.F. produced and developed the idea. S.P.F., Z.Y., A.Al. and T.A. carried out theoretical simulations of the system. L.C., E.M., T.L., S.H., A.Am., M.D. and M.F.G. contributed to the implementation of the control methods. S.K., M.R., A.O., A.F.R., M.C., L.G., and J.G. participated in the fabrication of the device. L.C., Z.Y., A.Al. wrote and G.T., A.F.K., and J.By. edited the manuscript. A.F.K., J.By., and G.T. provided supervision and guidance during the project. All authors contributed to the discussions and interpretations of the results.\par

\section*{COMPETING INTERESTS}
The authors declare no competing interests.\par

\bibliography{library}
\bibliographystyle{revtex.bst}

\clearpage
\newpage

\onecolumngrid

\section*{\NoCaseChange{Supplementary information for "Fast unconditional reset and\\ leakage reduction in fixed-frequency transmon qubits"}}

\twocolumngrid

\renewcommand{\theequation}{S\arabic{equation}}
\renewcommand{\thetable}{S\arabic{table}}
\renewcommand{\thefigure}{S\arabic{figure}}
\setcounter{equation}{0}
\setcounter{table}{0}
\setcounter{figure}{0}

\section*{\NoCaseChange{Supplementary Note I : Experimental setup}}
\label{sec_ap_expSet}

\begin{figure}[t!]
\centering
 \includegraphics [width=3.3in, keepaspectratio=true] {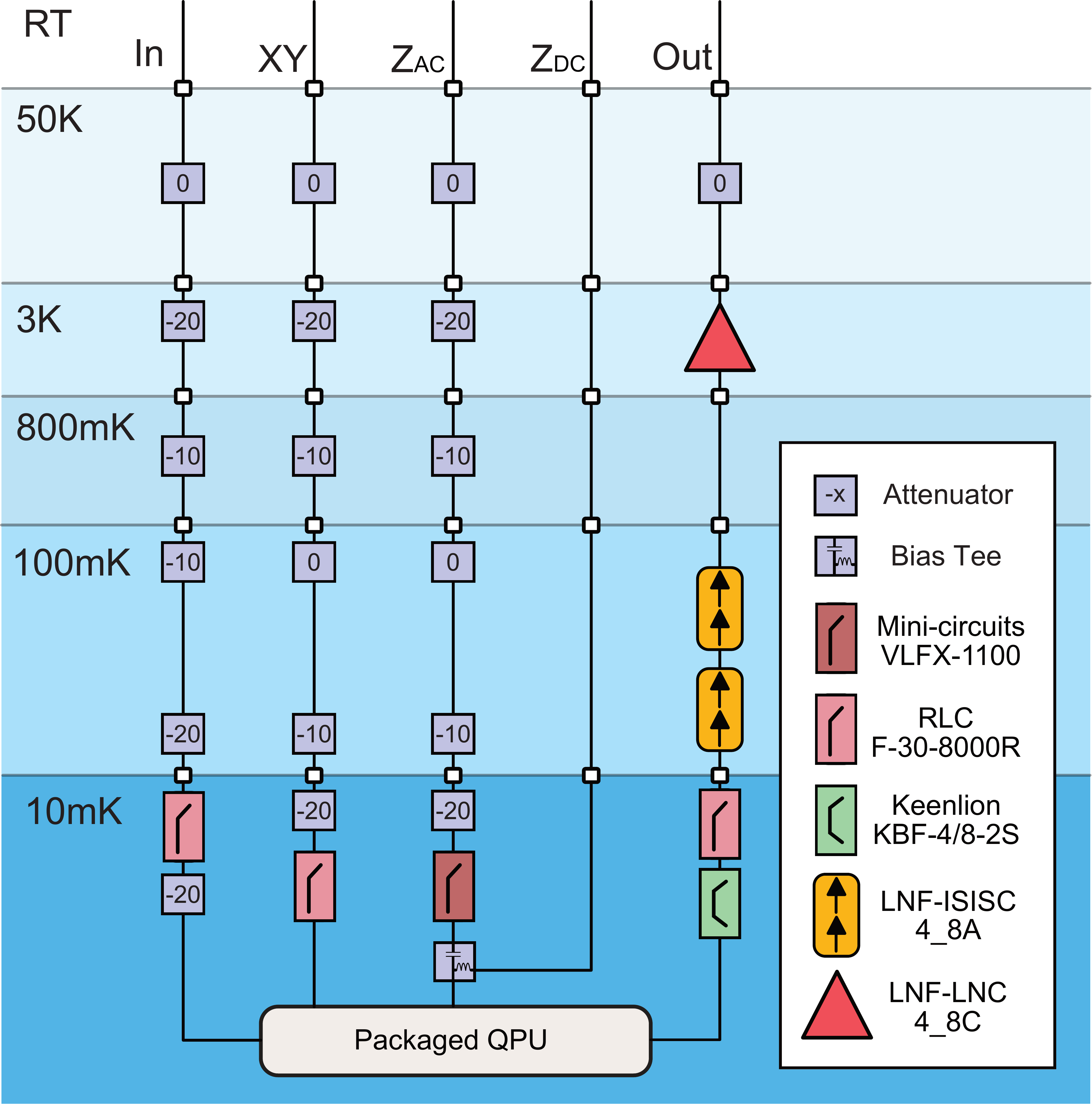}
\caption{\textbf{Full wiring diagram for the experimental setup including the circuit model of the tested chip.} XY: qubit XY-drive line, $\text{Z}_{DC}$: coupler bias current line, $\text{Z}_{AC}$: coupler flux pulse line, RT: room temperature.}
\label{fig_CablingTree}
\end{figure}

The parameters of the sample used in our experiment are listed in Table~\ref{Table: Para_singlequbit}. The device is installed in a Bluefors dilution refrigerator and cooled below 10~mK. The full wiring diagram is shown in \figref{fig_CablingTree}. The drive pulses for the qubits are generated by a two-channel signal generator ($QBLOX~Instruments$ QCM-RF).\par

The readout pulse consists of a flat-top pulse with Gaussian rising and falling edges with a width of $10\,$ns, which is generated by a high-frequency lock-in amplifier ($QBLOX~Instruments$ QRM-RF) to match the corresponding readout frequency of the individual resonators. The transmitted readout signal through the sample is amplified by around $39\,$dB at $4\,$K with a high-electron-mobility transistor (HEMT, LNF-LNC4\_8C) amplifier. Additionally, two amplifiers are connected to the readout chain outside of the dilution fridge. Finally, the output signal is down-converted to the hundreds of \unit{\mega\hertz} range to be recorded and integrated by the QRM-RF.\par

The biased current for the coupler is supplied by a current source ($QBLOX~Instruments$ S4g) and the flux pulse is generated by a four-channel arbitrary waveform generator ($QBLOX~Instruments$ QCM) operating at baseband frequency. The DC and AC flux signal is coupled together with a bias tee at  the 10\unit{\milli\kelvin} stage.\par

\begin{table}[h!]
\footnotesize
    \centering
    \begin{tabular}{|p{3.3cm} | c | c | c |} 
    \hline
        \multicolumn{2}{|c|}{Parameters} & $\text{Q}_{0}$ & $\text{Q}_{1}$  \\ 
        \hline
        Qubit frequency & $\omega_{q}/2\pi$ (\unit{\giga\hertz}) &  5.176	
 &  4.534  \\
        \hline
        Qubit anharmonicity & $\alpha/2\pi$ (\unit{\mega\hertz}) & -256 & -158  \\ 
        \hline

        Resonator frequency & $\omega_{r}/2\pi$ (\unit{\giga\hertz}) 
                         & 6.752	
 & 6.308 \\ 
        \hline
        Effective linewidth & $\kappa_{r}\left(\kappa_{f}\right)/2\pi$ (\unit{\mega\hertz}) &  0.427	
 &  0.294  \\ 
        \hline
        Qubit-readout coupling & $g/2\pi$ (\unit{\mega\hertz}) & 46	
 & 54 \\
        \hline
        Dispersive shift & $\chi/2\pi$ (\unit{\mega\hertz}) &  0.132	
 &  0.088  \\
        \hline
        
         Relaxation time & $\overline{T}_{1}$ ($\mu$s) & 30.6 & 83.8  \\
        \hline
         Decoherence time & $\overline{T}^*_{2}$ ($\mu$s) & 60.3 & 90.8  \\
        \hline
        Qubit-coupler coupling & $g/2\pi$ (\unit{\mega\hertz}) & 47	
 & 64 \\
 \hline
        Single-qubit gate fidelity & $\mathcal{F}_{1Q} (\%)$ & 99.6\% & 99.8\%  \\ 
        \hline
    \end{tabular}
\caption{\textbf{Measured qubit parameters, coherence properties, and single-qubit performance for the two qubits.} This pair of qubits is a subset of a larger 25-qubit device with similar performance.\label{Table: Para_singlequbit}}
\end{table}

\section*{\NoCaseChange{Supplementary Note II : Data qubit response}}
\label{sec_ap_data}

\begin{figure}[t!]
\centering
 \includegraphics [width=3.5in, keepaspectratio=true] {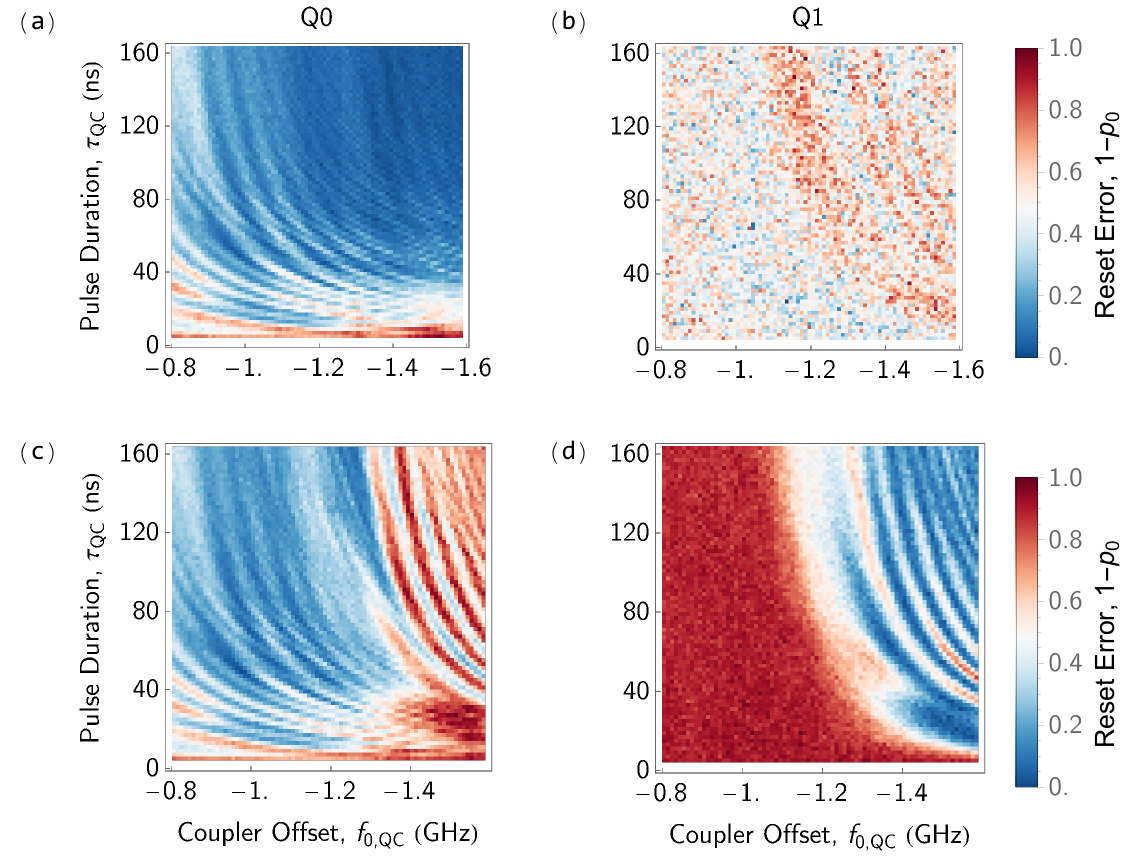}
\caption{\textbf{Adiabatic pulse performance with respect to the excitation on the data qubit.} In \textbf{(a-b)} (\textbf{(c-d)}), $\text{Q}_{1}$ is initially prepared in $\gstate$ ($\estate$)-state. The interaction between the coupler and $\text{Q}_{0}$ varies much differently depending on the state of $\text{Q}_{1}$. The coupling strength parameter $g$ is set to 1~\unit{\giga\hertz} for both measurements to approximate a linear ramp pulse. }
\label{fig_adibatic_data}
\end{figure}

One limiting constraint on the adiabaticity of the reset pulse is the frequency spacing between the two qubits. In \figref{fig_adibatic_data}, two sets of measurements are carried out for the same reset pulse with $\text{Q}_{1}$ initially prepared in either $\gstate$ [\figpanel{fig_adibatic_data}{a-b}] or $\estate$ [\figpanel{fig_adibatic_data}{c-d}]. We observe that the parameter space available for a complete reset for $\text{Q}_{0}$ in \figpanel{fig_adibatic_data}{a} is much larger than that in \figpanel{fig_adibatic_data}{c}. The transition in this regime is fully adiabatic, indicated by the fact that there is no oscillation occurring at a longer pulse duration.\par

However, reset pulses with these parameters move the coupler too close to the $\text{Q}_{1}$ frequency, resulting in significant undesirable interaction with $\text{Q}_{1}$, as shown more prominently in \figpanel{fig_adibatic_data}{d}. Therefore, in order to protect the $\estate$-state population of $\text{Q}_{1}$, we choose to operate in the regime indicated by \figpanel{fig_adibatic_data}{c}. This constraint increases the calibration difficulty and potentially decreases long-term stability in large devices.\par








\end{document}